\documentclass[12pt]{article}
\usepackage{amssymb}
\usepackage{amsmath}
\usepackage{amsfonts}

\newcommand{\R}{\mathbb{R}}
\newcommand{\C}{\mathbb{C}}

\newcommand{\be}{\begin{equation}}
\newcommand{\bea}{\begin{eqnarray}}
\newcommand{\eea}{\end{eqnarray}}
\newcommand{\nn}{\nonumber}
\newcommand{\kt}{\rangle}
\newcommand{\br}{\langle}

\newcommand{\ed}{\end{document}}

\begin{document}

\title{On the Representation Theory of Orthofermions and Orthosupersymmetric Realization of
Parasupersymmetry and Fractional Supersymmetry}
\author{Ali Mostafazadeh\thanks{E-mail address:
amostafazadeh@ku.edu.tr}\\ \\
Department of Mathematics, Ko\c{c} University,\\
Rumelifeneri Yolu, 80910 Sariyer, Istanbul, Turkey}
\date{ }
\maketitle

\begin{abstract}
We construct a canonical irreducible representation for the
orthofermion algebra of arbitrary order, and show that every
representation decomposes into irreducible representations that
are isomorphic to either the canonical representation or the
trivial representation. We use these results to show that every
orthosupersymmetric system of order $p$ has a parasupersymmetry of
order $p$ and a fractional supersymmetry of order $p+1$.
\end{abstract}

\baselineskip=24pt

\section{Introduction}

Orthofermions were originally introduced by Khare, Mishra, and Rajasekaran \cite{ortho} in an attempt
to obtain a generalization of supersymmetry called orthosupersymmetry. Recently \cite{p39}, it has
been realized that orthofermions may be used to construct parafermions of order 2, and that every
orthosupersymmetric system possesses topological symmetries \cite{p34b}. In particular, given an
orthosupersymmetric system of order $p$, one can construct a fractional supersymmetric system of
order $p+1$ \cite{p39}. The main ingredient leading to these observations is the algebra of
orthosupersymmetric quantum mechanics \cite{ortho}:
    \bea
    &&[H,Q_\alpha]=0\;,
    \label{os1}\\
    &&Q_\alpha Q_\beta^\dagger +\delta_{\alpha \beta} \sum^p_{\gamma=1}Q_
    \gamma^\dagger Q_\gamma=2\delta_{\alpha \beta}H\;,
    \label{os2}\\
    &&Q_\alpha Q_\beta=0\;,
    \label{os3}
    \eea
where $Q_\alpha$ are the generators of the orthosupersymmetry, $\alpha,\beta\in\{1,2,\cdots,p\}$, and
$\delta_{\alpha\beta}$ stands for the Kronecker delta function. The simplest quantum system
possessing orthosupersymmetry of order $p$ is a system with bosonic and orthofermionic degrees of
freedom. For this system the generators of orthosupersymmetry are given by
$Q_\alpha={\sqrt 2}a^\dagger c_\alpha$ where $a$ is the annihilation operator for a boson and
$c_\alpha$ are the annihilation operators for orthofermions of order $p$. They are defined through the
relations \cite{ortho}
    \bea
    &&[a,a^\dagger]=1,~~~~[a,c_\alpha]=[a,c_\alpha^\dagger]=0,\nn\\
    &&c_\alpha c_\beta^\dagger +\delta_{\alpha \beta}
    \sum^p_{\gamma=1}c_\gamma^\dagger c_\gamma=\delta_{\alpha \beta} 1\;,
    \label{of1}\\
    &&c_\alpha c_\beta=0\;,
    \label{of2}
    \eea
where 1 stands for the identity operator. The study of orthosupersymmetry \cite{ortho} relies on a
matrix representation of orthofermions of order $p$ where $c_\alpha$ are represented by
$(p+1)\times(p+1)$ matrices with entries
    \be
    [c_\alpha]_{ij}=\delta_{i,1}\delta_{j,\alpha+1}\;,~~~ \forall i,j\in\{1,\cdots,p+1\}\;.
    \label{rep}
    \end{equation}
The purpose of this article is to study the general representations of the orthofermion algebra, i.e.,
Eqs.~(\ref{of1}) and (\ref{of2}), and to explore the implications of this study for ortho-, para-, and
fractional supersymmetry of arbitrary order.

The organization of the article is as follows. In Section~2, we
construct a canonical irreducible representation for the
orthofermion algebra. In Section~3, we examine general
representations of the orthofermion algebra, and show that every
representation decomposes into the irreducible representations
that are either isomorphic to the canonical representation or the
trivial representation. In Section~4, we construct the ladder
operators for the canonical representation and derive some of
their basic properties. In Section~5, we use the results of the
preceding sections to show that every orthosupersymmetric system
of order $p$ possesses a parasupersymmetry of order $p$ and a
fractional supersymmetry of order $p+1$. In Sections~6, we
summarize our results and present our concluding remarks.

\section{The Canonical Irreducible Representation of the Orthofermion Algebra}

We begin our analysis of the orthofermion algebra, i.e., Eqs.~(\ref{of1}) and (\ref{of2}), by
introducing
    \be
    \Pi:=1-\sum_{\alpha=1}^p c_\alpha^\dagger c_\alpha\;.
    \label{pi}
    \end{equation}
This allows us to write Eq.~(\ref{of1}) in the form
    \be
    c_\alpha c_\beta^\dagger=\delta_{\alpha\beta} \Pi\;.
    \label{of3}
    \end{equation}
It is not difficult to show that $\Pi$ is a Hermitian projection operator:
    \be
    \Pi^2=\Pi=\Pi^\dagger.
    \label{pi-1}
    \end{equation}
This follows from Eqs.~(\ref{of2}), (\ref{pi}), and (\ref{of3}). Furthermore, for all $\alpha\in\{1,2,\cdots,p\}$,
    \bea
    \Pi\, c_\alpha&=&c_\alpha,~~~c_\alpha^\dagger\Pi=c_\alpha^\dagger,
    \label{pi-2}\\
    c_\alpha\Pi&=&0,~~~\Pi\, c_\alpha^\dagger=0.
    \label{pi-3}
    \eea

Next, let ${\cal A}$ denote the (complex associative $*$) algebra generated by $2p$ generators:
$c_\alpha,c_\alpha^\dagger$ with $\alpha\in\{1,2,\cdots,p\}$, and subject to relations (\ref{of2}),
(\ref{pi}), and (\ref{of3}).\footnote{Clearly, the $*$ operation is given by $\dagger$.} Then in view of
these relations and Eqs.~(\ref{pi-2}) and (\ref{pi-3}), elements of ${\cal A}$ have the general form
    \be
    x=\lambda\Pi+\sum_{\alpha=1}^p(\nu_\alpha c_\alpha +\mu_\alpha c_\alpha^\dagger)+
    \sum_{\alpha,\beta=1}^p\sigma_{\alpha\beta}\,c_\alpha^\dagger c_\beta,
    \label{x=}
    \end{equation}
where $\lambda,\nu_\alpha,\mu_\alpha$, and $\sigma_{\alpha\beta}$ are complex numbers. As seen
from Eq.~(\ref{x=}), ${\cal A}$ is a $(p+1)^2$-dimensional complex vector space. We can use this
vector space as a representation space for orthofermion algebra. However, as we shall see in
Section~4 this would lead to a reducible representation. Therefore, we will restrict to a
subrepresentation.

Let ${\cal A}_0\subset{\cal A}$ be the span of $\Pi$ and $c_\alpha^\dagger$, and
    \[x_0:=\lambda \Pi+\sum_{\alpha=1}^p\mu_\alpha c_\alpha^\dagger\]
be an arbitrary element of ${\cal A}_0$. Then, in view of Eq.~(\ref{of3}) and (\ref{pi-2}), for all $\alpha\in\{1,2,\cdots,p\}$,
    \be
    c_\alpha x_0= \mu_\alpha \Pi,~~~~~c_\alpha^\dagger x_0=\lambda c_\alpha^\dagger.
    \label{fock}
    \end{equation}
These equations suggest that ${\cal A}_0$ is the representation space for a Fock space representation
of orthofermions. Following the standard notation, we set
    \be
    |0\kt:=\Pi,~~~{\rm and}~~~\forall\alpha\in\{1,2,\cdots,p\},~~~|\alpha\kt:=c_\alpha^\dagger.
    \label{fock-1}
    \end{equation}
This yields
    \be
    \forall\alpha\in\{1,2,\cdots,p\},~~~c_\alpha|0\kt=0~~~{\rm and}~~~|\alpha\kt=
    c_\alpha^\dagger|0\kt.
    \label{fock-2}
    \end{equation}
Furthermore, using Eq.~(\ref{pi-3}) we have
    \be
    \Pi|0\kt=|0\kt,~~~{\rm and}~~~\forall\alpha\in\{1,2,\cdots,p\},~~~\Pi|\alpha\kt=0.
    \label{pi-4}
    \end{equation}
Therefore, $\Pi$ is the projection onto the `vacuum' state vector $|0\kt$.

As a vector space ${\cal A}_0$ is isomorphic to $\C^{p+1}$. The vectors $|n\kt$ with
$n\in\{0,1,\cdots,p\}$ form a basis for ${\cal A}_0$. In this basis, $|n\kt$ may be identified with
column vectors whose $k$-th component is given by $\delta_{nk}$ and the operators $c_\alpha$ are
represented by $(p+1)\times (p+1)$ matrices with entries
    \be
    [c_\alpha]_{ij}=\delta_{i,1}\delta_{j,\alpha+1}.
    \label{irrep}
    \end{equation}
This is precisely the matrix representation (\ref{rep}) of Ref.~\cite{ortho}. Note that unlike in
Ref.~\cite{ortho}, here we construct the representation space: ${\cal A}_0$. Since, we have obtained
the action of $c_\alpha$ and $c_\alpha^\dagger$ on the basis vectors $|n\kt$, we can represent all the
elements of ${\cal A}$ by linear operators (endomorphisms) mapping ${\cal A}_0$ into itself, i.e., we
have a representation $\rho_0:{\cal A}\to{\rm End}({\cal A}_0)$ of the algebra ${\cal A}$. Here
`End' abbreviates the `space of endomorphisms of,' and by a representation
$\rho:{\cal A}\to{\rm End}(V)$ in a complex vector space $V$ we mean a linear map satisfying
    \be
    \rho(x_1x_2)=\rho(x_1)\rho(x_2),~~~\forall x_1,x_2\in {\cal A}.
    \label{rep1}
    \end{equation}
We shall also postulate
    \be
    \forall x\in {\cal A},~~~\rho(x^\dagger)=\rho(x)^\dagger,
    \label{rep2}
    \end{equation}
if $V$ is endowed with an inner product.

Note that the representation $\rho_0$ is an irreducible representation. This may be easily verified by
inspecting the matrices~(\ref{irrep}).

\section{Representation Theory of the Orthofermion Algebra}

Let $V$ be an arbitrary complex vector space, $\rho:{\cal A}\to{\rm End}(V)$ be a representation of
${\cal A}$, and $V_0$ be the subspace of $V$ defined by
    \[ V_0:={\rm Im}(\rho(\Pi)):=\left\{ \rho(\Pi)v~|~v\in V\right\}.\]
\begin{itemize}
\item[~]{\bf Lemma~1:} If $V_0=\{0\}$, then $\rho$ is a trivial representation, i.e., for all $x\in {\cal A}$ and
$v\in V$, $\rho(x)v=0$.
\item[~]{\bf Proof:} In view of Eq.~(\ref{rep1}), it is sufficient to prove that $\rho(c_\alpha)v=\rho(c_\alpha^\dagger)v=0$.
Because $V_0=\{0\}$, for all $u\in V$, $\rho(\Pi)u=0$. But then according to Eqs.~(\ref{rep1}), (\ref{pi-2}) and (\ref{pi-3}),
    \[ \rho(c_\alpha)v=\rho(\Pi\,c_\alpha)v=\rho(\Pi)[\rho(c_\alpha)v]=0,~~~~
    \rho(c_\alpha^\dagger)v=\rho(c_\alpha^\dagger\Pi)v=\rho(c_\alpha^\dagger)[\rho(\Pi)v]=0.~~~\square\]
\end{itemize}
Now, suppose that $\rho$ is not a trivial representation. Then $V_0$ is a nontrivial subspace of $V$. Let $\{e_i\}$ be a
basis of $V_0$ and $V_i$ be the subspaces of $V$ defined by
    \[V_i:={\rm Span}\left( \left\{e_{i}\right\} \bigcup
    \left\{\rho(c_\alpha^\dagger)e_{i}\right\}_{\alpha\in\{1,2,\cdots,p\}}\right).\]
Then we can prove the following lemma.
\begin{itemize}
\item[~]{\bf Lemma~2:} For all $x\in {\cal A}$, $\rho(x)$ maps $V_i$ into $V_i$.
\item[~]{\bf Proof:} It suffices to show that for all $\alpha\in\{1,2,\cdots,p\}$, $\rho(c_\alpha)$ and
$\rho(c_\alpha^\dagger)$ map $V_i$ into $V_i$. Let $v\in V_i$, then there are complex numbers $\mu_\alpha$ and
$\lambda$ such that $v=\sum_{\beta=1}^p\mu_\beta \rho(c_\beta^\dagger)e_i+\lambda e_i$. In view of
Eqs.~(\ref{rep1}), (\ref{pi-3}), (\ref{pi-1}), and $e_{i_\ell}\in V_\ell$,
    \bea
    \rho(c_\alpha)v&=&\sum_{\alpha=1}^p\mu_\beta\rho(c_\alpha c_\beta^\dagger)e_i+\lambda\rho(c_\alpha)e_i
    =\mu_\alpha e_i\in V_i\nn\\
    \rho(c_\alpha^\dagger)v&=&\sum_{\alpha=1}^p\mu_\beta\rho(c_\alpha^\dagger c_\beta^\dagger)e_i+
    \lambda\rho(c_\alpha^\dagger)e_i=\lambda\rho(c_\alpha^\dagger)e_i\in V_i.~~~\square\nn
    \eea
\end{itemize}
A direct implication of Lemma~2 is that for each basis vector $e_i$ of $V_0$ the restriction
    \[\rho_i=\left.\rho\right|_{V_i}:{\cal A}\to {\rm End}(V_i)\]
of $\rho$ provides a representation of the algebra ${\cal A}$.

Furthermore, introducing
    \[|0\kt_i:=e_{i},~~~{\rm and}~~~|\alpha\kt_i:=\rho_i(c_\alpha^\dagger)|0\kt_i,~~~\forall\alpha\in\{1,2,\cdots,p\},\]
we can easily show that $|n\kt_i$, with $n\in\{0,1,\cdots,p\}$, are basis vectors for $V_i$ and that in this basis the
operators $\rho_i(c_\alpha)$ are represented by matrices whose entries are given by the right-hand side of Eq.~(\ref{rep}).
Therefore, the representations $\rho_i$ are equivalent to the representation $\rho_0$. In particular, they are irreducible
representations.

Next, consider the case where the dimension of $V_0$ is greater than one.
\begin{itemize}
\item[~]{\bf Lemma~3:} Let $e_{i_1}$ and $e_{i_2}$ be distinct basis vectors of $V_0$. Then $V_{i_1}\cap V_{i_2}=\{0\}$.
\item[~]{\bf Proof:} Suppose $v\in V_{i_1}\cap V_{i_2}$. Then there are complex numbers $\mu_\alpha,\mu'_\alpha,
\lambda$, and $\lambda'$ such that
    \be
    v=\sum_{\alpha=1}^p\mu_\alpha \rho(c_\alpha^\dagger)e_{i_1}+\lambda e_{i_1}=
    \sum_{\alpha=1}^p\mu'_\alpha \rho(c_\alpha^\dagger)e_{i_2}+\lambda' e_{i_2}.
    \label{v=}
    \end{equation}
Applying $\rho(\Pi)$ to both sides of the second equation
in~(\ref{v=}) and using (\ref{rep1}), (\ref{pi-3}), (\ref{pi-1}),
and $e_{i_\ell}\in V_\ell$, we find $\lambda e_{i_1}=\lambda'
e_{i_2}$, which implies $\lambda=\lambda'=0$. Similarly, applying
$\rho(c_\beta)$ to both sides of (\ref{v=}) for an arbitrary
$\beta\in\{1,2\cdots,p\}$, we have $\mu_\beta e_{i_1}=\mu'_\beta
e_{i_2}$ which yields $\mu_\beta=\mu'_\beta=0$. Hence,
$v=0$.~~~$\square$
\end{itemize}

Now, we are in a position to address the issue of the decomposition of an arbitrary representation $\rho$ into irreducible
representations. The algebra ${\cal A}$ does not contain a unit~1. In the following we shall extend ${\cal A}$ by
adding $1$ as a generator satisfying: $\forall x\in {\cal A}, 1x=x1=x$. Inclusion of~1 allows us to use Eq.~(\ref{pi}) in the
representations of ${\cal A}$. Clearly, we have $\forall x \in{\cal A}, \rho(1)\rho(x)=\rho(x)\rho(1)=\rho(x)$.
For the representations $\rho_i$, we have $\rho_i(1)=I_i$, where $I_i$ is the identity operator acting on $V_i$. This follows
from the equivalence of $\rho_i$ and $\rho_0$. Note also that for a trivial representation, we have $\rho(1)=0$.
\begin{itemize}
\item[~]{\bf Lemma~4:} Suppose that $\rho(1)=I$, where $I$ denotes the identity operator acting on $V$, and let
$V_*:=\oplus_i V_i$. Then $V=V_*$.
\item[~]{\bf Proof:} Let $w\in V-V_*$, in particular $w\neq 0$, and $e:=\rho(\Pi)w\in V_0\subset V_*$, so that $e\in V_*$
and $e\neq w$. Now, using Eq.~(\ref{pi}), we have
    \[\sum_{\alpha=1}^p\rho(c_\alpha^\dagger)\rho(c_\alpha)w=\rho(1)w-e=w-e\notin V_*.\]
This in turn implies that $\rho(c_\alpha) w\notin V_*$. Also in view of Eq.~(\ref{pi-2}), we have
$\rho(\Pi)\rho(c_\alpha)w=\rho(c_\alpha) w\notin V_*$. This contradicts $\rho(\Pi)(\rho(c_\alpha)w)\in V_0\subset V_*$.
Therefore, such a $w$ does not exist and $V=V_*$.~~~$\square$
\end{itemize}

Next consider the case where $V$ is endowed with an inner product $\br~|~\kt$. Then we have the following results.
\begin{itemize}
\item[~]{\bf Lemma~5:}  Let $e_{i_1}$ and $e_{i_2}$ be orthogonal basis vectors of $V_0$. Then $V_{i_1}$ and $V_{i_2}$ are
orthogonal subspaces of $V$.
\item[~]{\bf Proof:} This statement follows from the identities
    \bea
    \br \rho(c_\alpha^\dagger)e_{i_1}|\rho(c_\beta^\dagger)e_{i_2}\kt&=&
    \br e_{i_1}|\rho(c_\alpha c_\beta^\dagger)e_{i_2}\kt
    =\delta_{\alpha\beta}\br e_{i_1}|\rho(\Pi)|e_{i_2}\kt
    =\delta_{\alpha\beta}\br e_{i_1}|e_{i_2}\kt=0,\nn\\
    \br e_{i_1}|\rho(c_\alpha^\dagger)e_{i_2}\kt&=&\br\rho(c_\alpha)e_{i_1}|e_{i_2}\kt=0,\nn
    \eea
where we have made use of Eq.~(\ref{rep2}) and $\rho(c_\alpha)e_{i_1}=0$.~~~$\square$
\end{itemize}
This lemma implies that $V_*$ is actually an orthogonal direct sum of the subspaces $V_i$.
\begin{itemize}
\item[~]{\bf Lemma~6:} Let $V_*^c:=\{w\in V|\forall v\in V_*, \br w|v\kt=0\}$ be the orthogonal complement of $V_*$. Then
$V_*^c$ is the representation space for a trivial representation of ${\cal A}$.
\item[~]{\bf Proof:} Let $w\in V_*^c$ and $e:=\rho(\Pi)w\in V_0\subset V_*$. Then clearly, $\br w|e\kt=0$ and
    \be
    \br e|e\kt=\br \rho(\Pi)w|\rho(\Pi)w\kt=\br w|\rho(\Pi)^\dagger\rho(\Pi)w\kt=\br w|\rho(\Pi)w\kt=\br w|e\kt=0.
    \label{ee}
    \end{equation}
Here, we have made use of Eqs.~(\ref{pi-1}) and (\ref{rep2}). Eq.~(\ref{ee}) implies $\rho(\Pi)w=e=0$. Hence $\rho(\Pi)(V_*^c)=\{0\}$.
Now, computing
    \[ \br\rho(c_\alpha^\dagger)w|\rho(c_\alpha^\dagger)w\kt=\br w|\rho(c_\alpha)\rho(c_\alpha^\dagger)w\kt=
    \br w|\rho(\Pi)w\kt=0,\]
we find that
    \be
    \rho(c_\alpha^\dagger)w=0.
    \label{cw=0}
    \end{equation}
Next, using Eq.~(\ref{pi}), we have
    \be
    [\rho(1)-\sum_{\alpha=1}^p\rho(c_\alpha^\dagger)\rho(c_\alpha)]w=\rho(\Pi)w=0.
    \label{id-1}
    \end{equation}
Furthermore, let us express $\rho(c_\alpha)w=w_\alpha+w_\alpha^c$ where $w_\alpha\in V_*$ and $w_\alpha^c\in V_*^c$.
Then according to the above argument $\rho(c_\alpha^\dagger)w_\alpha^c=0$. This together with Eqs.~(\ref{cw=0}) and (\ref{id-1})
imply
    \be
    \br\rho(1)w|\rho(1)w\kt=\br w|\rho(1^\dagger)\rho(1)w\kt=
    \br w|\rho(1)w\kt=\sum_{\alpha=1}^p\br w|\rho(c_\alpha^\dagger)\rho(c_\alpha)w\kt=
    \sum_{\alpha=1}^p\br w|\rho(c_\alpha^\dagger)w_\alpha\kt=0.
    \label{ww}
    \end{equation}
The last equality follows from the fact that since $w_\alpha\in V_*$, $\rho(c_\alpha^\dagger)w_\alpha\in V_*$. Eq.~(\ref{ww})
implies $\rho(1)w=0$. Therefore, for all $x\in A$,
    \[ \rho(x)w=\rho(x1)w=\rho(x)\rho(1)w=0,\]
and $V_*^c$ yields a trivial representation of ${\cal A}$.~~~$\square$
\end{itemize}

In summary, up to equivalence, the orthofermion algebra of order $p$ has a unique nontrivial $(p+1)$-dimensional irreducible
representation $\rho_0$, and every representation decomposes into irreducible representations that are equivalent to either
the trivial representation or $\rho_0$. In particular, the orthosupersymmetry algebra is in a sense the unique generalization
of the supersymmetry algebra describing bose-orthofermi symmetry.

\section{The Ladder Operators of the Canonical Representation}

Consider the canonical irreducible representation $\rho_0$ of Section~2 and let
    \be
    L:=c_1+\sum_{\alpha=2}^p c_{\alpha-1}^\dagger c_{\alpha}\;.
    \label{L}
    \end{equation}
Then, in view of Eqs.~(\ref{of3}), (\ref{fock-1}) -- (\ref{pi-4}), we have
    \be
    L|n\kt=\left\{ \begin{array}{ccc}
    0&{\rm for}& n=0\\
    |n-1\kt&{\rm for}&n\in\{1,2,\cdots,p\}\end{array}\right.
    ~~~~~
    L^\dagger|n\kt=\left\{ \begin{array}{ccc}
    |n+1\kt&{\rm for}& n\in\{0,1,\cdots,p-1\}\\
    0&{\rm for}&n=p\end{array}\right.
    \label{ladder}
    \end{equation}
These equations show that $L$ and $L^\dagger$ are the ladder operators for the canonical
representation of the orthofermion algebra.

The ladder operators $L$ and $L^\dagger$ have certain interesting properties. For example,  we can
use Eqs.~(\ref{L}), (\ref{of3}), (\ref{pi-2}), and (\ref{pi-3}), to compute
    \bea
    L^\dagger L&=&1-\Pi\;,
    \label{LL1}\\
    LL^\dagger&=&1-c_p^\dagger c_p\;,
    \label{LL2}\\
    L^k&=&\left\{\begin{array}{ccc}
    c_k+\sum_{\alpha=1}^{p-k}c_\alpha^\dagger c_{\alpha+k}&{\rm for}&k\in\{1,2,\cdots,p-1\}\\
    c_p&{\rm for}& k=p\\
    0&{\rm for}& k\in\{p+1,p+2,\cdots\},\end{array}\right.
    \label{L1}\\
    L^pL^\dagger&=&c_{p-1}\,,~~~~L^\dagger L^p=L^{p-1}-c_{p-1}\;.
    \label{L2}
    \eea
In view of these equations and Eqs.~(\ref{pi-2}), and (\ref{pi-3}), we also obtain
    \bea
    &&L^k\Pi=0\,,~~~~\Pi L^k=c_k\,,
    \label{L3}\\
    &&L^{p-k}L^\dagger L^k=L^{p-1}\;,
    \label{L4}
    \eea
where $k\in\{1,2,\cdots,p\}$.

Eqs.~(\ref{L1}) -- (\ref{L4}) imply
    \bea
    &&L^{p+1}=0\;,
    \label{L-para-1}\\
    &&\sum_{k=0}^p L^{p-k} L^\dagger L^k=p\, L^{p-1}\;.
    \label{L-para-2}
    \eea
These equations are reminiscent of the defining equations for the
parasupersymmetry of order $p$, \cite{psusy}. In section~5, We
shall use these equations to establish that every
orthosupersymmetric system of order $p$ has a parasupersymmetry of
order $p$.

Next, let
    \be
    F:=L+c_p^\dagger.
    \label{F}
    \end{equation}
Then, in view of Eqs.~(\ref{fock-1}), (\ref{fock-2}), and (\ref{ladder}), we have
    \be
    F|n\kt=\left\{\begin{array}{ccc}
    |n-1\kt &{\rm for}& n\in\{1,\cdots,p\}\\
    |p\kt &{\rm for}&n=0,\end{array}\right.~~~
    F^\dagger|n\kt=\left\{\begin{array}{ccc}
    |n+1\kt &{\rm for}& n\in\{0,1,\cdots,p-1\}\\
    |0\kt &{\rm for}&n=p.\end{array}\right.
    \label{cycle}
    \end{equation}
In particular,
    \be
    F^{p+1}=1\;.
    \label{F1}
    \end{equation}
In the following section, we shall make use of this identity to show that every orthosupersymmetric
system of order $p$ has a fractional supersymmetry of order $p+1$.

\section{An Orthosupersymmetric Realization of Parasupersymmetry and Fractional Supersymmetry}

In Ref.~\cite{p39}, the algebra (\ref{os1}) -- (\ref{os3}) is used to show that the operator
    \[\tilde Q:=Q_1^\dagger +\sum_{\alpha=2}^p Q_{\alpha}^\dagger Q_{\alpha-1}+Q_p\]
satisfies $\tilde Q^{p+1}=(2H)^p$. Therefore, $\tilde Q$ is the
generator of a fractional supersymmetry for the Hamiltonian
$K:=(2H)^p$. In this section, we shall demonstrate that any
orthosupersymmetric system has a fractional supersymmetry of order
$p+1$ and a parasupersymmetry of order $p$.

First, we recall that the energy spectrum of an orthosupersymmetric Hamiltonian $H$ is nonnegative.
This follows from Eq.~(\ref{os2}). Setting $\alpha=\beta$ in this equation, we have for any state
vector $|\psi\kt$,
    \be
    \br\psi|H|\psi\kt=
    ||Q_\alpha^\dagger|\psi\kt ||^2+\sum_{\gamma=1}^p||Q_\gamma|\psi\kt||^2\geq 0.
    \label{<H>}
    \end{equation}
Next, Let $E$ denote an eigenvalue of $H$, and ${\cal H}^{(E)}$ denote the corresponding
eigenspace. Because of Eq.~(\ref{os1}) the restriction $Q_1^{(E)}:=Q|_{{\cal H}^{(E)}}$ is an
operator mapping ${\cal H}^{(E)}$ into ${\cal H}^{(E)}$. Restricting Eqs.~(\ref{os2}) and
(\ref{os3}) to ${\cal H}^{(E)}$, we find
    \bea
    &&Q_\alpha^{(E)} Q_\beta^{(E)\dagger} +\delta_{\alpha \beta}
    \sum^p_{\gamma=1}Q_\gamma^{(E)\dagger} Q_\gamma^{(E)}=
    2\delta_{\alpha \beta}E~I^{(E)} \;,
    \label{os2-E}\\
    &&Q_\alpha^{(E)} Q_\beta^{(E)}=0\;,
    \label{os3-E}
    \eea
where $I^{(E)}$ denotes the identity operator on ${\cal H}^{(E)}$.

Now, if $E=0$, then according to Eq.~(\ref{<H>}), we have
    \be
    Q_\alpha^{(0)}=Q_\alpha^{(0)\dagger}=0\;.
    \label{trivial}
    \end{equation}
Next, introduce
    \be
    c_\alpha^{(E)} :=\left\{\begin{array}{ccc}
    0&{\rm for}&E=0\\
    (2E)^{-1/2}Q_\alpha^{(E)}&{\rm for}&E>0.
    \end{array}\right.
    \label{q-rep}
    \end{equation}
Then in terms of $c_\alpha^{(E)} $, Eqs.~(\ref{os2-E}) and (\ref{os3-E}), take the form
    \bea
    &&c_\alpha^{(E)} c_\beta^{(E)^\dagger}+\delta_{\alpha \beta}
    \sum^p_{\gamma=1}c_\gamma^{(E)\dagger} c_\gamma^{(E)}=\left\{\begin{array}{ccc}
    0&{\rm for}& E=0\\
    2\delta_{\alpha \beta} I^{(E)}&{\rm for}&E>0\;,\end{array}\right.
    \label{os2-c}\\
    &&c_\alpha^{(E)} c_\beta^{(E)} =0\;.
    \label{os3-c}
    \eea
Comparing these equations with Eqs.~(\ref{of1}) and (\ref{of2}), we see that $c_\alpha^{(E)}$
provide a representation $\rho^{(E)}$ of the orthofermion algebra,
    \be
    c_\alpha^{(E)}=\rho^{(E)}(c_\alpha).
    \label{c=c}
    \end{equation}
Clearly, for $E=0$, this representation is the direct sum of a number
$n_0={\rm dim}({\cal H}^{(0)})$ of trivial representations. Moreover, in view of Eq.~(\ref{os2-c}),
for $E>0$, the identity operator $1$ is represented by $I^{(E)}$. Therefore, according to Lemmas~4
and~5, the representation $\rho^{(E)}$ decomposes into a number $n_E$ of irreducible
representations $\rho_i^{(E)}$ which are equivalent to the canonical representation $\rho_0$. Denoting
the corresponding representation spaces by ${\cal H}_i^{(E)}$, we can express ${\cal H}^{(E)}$ as
an orthogonal direct sum of ${\cal H}^{(E)}_i$,
    \[{\cal H}^{(E)}=\oplus_{i=1}^{n_E}{\cal H}^{(E)}_i\]

A direct implication of the fact that $c_\alpha^{(E)}$ provide a representation $\rho^{(E)}$ which in
turn decomposes into the irreducible representations $\rho_i^{(E)}$ is that the positive energy
eigenvalues $E$ are $n_i(p+1)$-fold degenerate. This confirms the results of Ref.~\cite{p39}
on the topological symmetries \cite{p34b} of orthosupersymmetric systems.

Next, let $L^{(E)}:=\rho^{(E)}(L)$ and $F^{(E)}:=\rho^{(E)}(F)$, where $L$ and $F$ are the
operators introduced in Eqs.~(\ref{L}) and (\ref{F}), respectively. In view of the above mentioned
decomposition of $\rho^{(E)}$ into $\rho^{(E)}_i$, the equivalence of the latter with $\rho_0$, and
Eqs.~(\ref{L-para-1}), (\ref{L-para-2}) and (\ref{F1}), we have
    \bea
    (L^{(E)})^{p+1}&=&0\;,~~~~~\sum_{K=0}^p(L^{(E)})^{p-k}
    L^{(E)\dagger} (L^{(E)})^k=p\, (L^{(E)})^{p-1}\;,
    \label{L-para}\\
    (F^{(E)})^{p+1}&=&1\;.
    \label{F-frac}
    \eea
Now, consider the operators $Q$ and ${\cal Q}$ defined through their restrictions $Q^{(E)}$ and
${\cal Q}^{(E)}$ on the eigenspaces ${\cal H}^{(E)}$ according to
    \bea
    Q^{(E)}&:=&\left\{\begin{array}{ccc}
        0 &{\rm for }& E=0\\
        \sqrt{2E}\,L^{(E)}&{\rm for}& E>0.\end{array}\right.
    \label{Q=}\\
    {\cal Q}^{(E)}&:=&\left\{\begin{array}{ccc}
        0 &{\rm for }& E=0\\
        E^{1/(p+1)}F^{(E)}&{\rm for}& E>0.\end{array}\right.
    \label{cuq=}
    \eea
Then, in view of Eqs.~(\ref{L-para}) and (\ref{F-frac}), we have
    \bea
    Q^{p+1}&=&0,~~~~\sum_{k=0}^p Q^{p-k}Q^\dagger Q^k=p\, Q^{p-1}H\;,
    \label{para-Q}\\
    {\cal Q}^{p+1}&=&H\;.
    \label{frac-cuq}
    \eea
Furthermore, by construction,
    \[ [Q,H]=[{\cal Q},H]=0.\]
These equations indicate that the system has a parasupersymmetry
\cite{psusy} of order $p$ generated by $Q$ and a fractional
supersymmetry \cite{fsusy} of order $p+1$ generated by ${\cal Q}$.

Note also that the parasupersymmetry and fractional supersymmetry generators can be expressed in
terms of the orthosupersymmetry generators $Q_\alpha$ according to
    \bea
    Q&=&Q_1+(2H)^{-1/2}\sum_{\alpha=2}^p Q_{\alpha-1}^\dagger Q_\alpha\;,\nn\\
    {\cal Q}&=&2^{-1/2} H^{-\frac{p-1}{p+1}} Q_1+2^{-1}H^{-\frac{p}{p+1}}
    \sum_{\alpha=2}^pQ_{\alpha-1}^\dagger Q_\alpha+2^{-1/2} H^{-\frac{p-1}{p+1}}
    Q_p^\dagger.\nn
    \eea
Here, for all $a\in\R^+$, $H^a:=\sum_E E^a \Lambda_E$, and $\Lambda_E$ is the projection
operator onto the eigenspace ${\cal H}^{(E)}$.

\section{Summary and Conclusion}

In this article, we addressed the representation theory of orthofermions. We constructed a canonical
$(p+1)$-dimensional irreducible representation for the orthofermion algebra of order $p$, and showed
that every representation of this algebra decomposes into copies of the trivial and the canonical
representation. The canonical representation which is a Fock space representation admits ladder
operators. We obtained these ladder operators and their properties to establish parasupersymmetry
and fractional supersymmetry of general orthosupersymmetric Hamiltonians. Our results may be
viewed as a novel realization of parasupersymmetry and fractional supersymmetry of arbitrary order. In
a sense, it yields an alternative statistical interpretation of these symmetries.

As argued in Ref.~\cite{p39} and shown in this paper, orthosupersymmetric systems satisfy the
defining properties of certain topological symmetries \cite{p34b}. The latter are a class of
generalizations of supersymmetry that involve topological invariants similar to the Witten index. A
proper understanding of these invariants requires the study of concrete toy models displaying these
symmetries. The orthosupersymmetric systems provide a class of these models. Our analysis of
orthofermion algebra leads to a clear picture of the general properties of orthosupersymmetry in
one dimension. A logical extension of our results would be to treat orthofermions and orthosupersymmetry
in higher dimensions. This might also shed some light on fractional supersymmetry in higher dimensions.

\section*{Acknowlegment}
I wish to thank Keivan Aghababaei Samani for reading the first
draft of the paper and for his comments. This project was
supported by the Young Researcher Award Program (GEBIP) of the
Turkish Academy of Sciences.


\ed
\begin{thebibliography}{9}
\bibitem{ortho} A.~Khare, A.~K.~Mishra and G.~Rajasekaran, Int.\
J.\ Mod.\ Phys.\ A {\bf 8}, 1245 (1993).
\bibitem{p39}  K.~Aghababaei Samani and A.~Mostafazadeh, `On the statistical origin of
topological symmetries,' hep-th/0105013.
\bibitem{p34b}  K.~Aghababaei Samani and A.~Mostafazadeh , Nucl.\
Phys.\ B {\bf 595}, 467 (2001).
\bibitem{psusy} A.~Khare, J.~Math.\ Phys.\ {\bf 34}, 1277 (1993); See also\\
M.~Tomiya, J.~Phys.\ A: Math.\ Gen.\ {\bf 25}, 4699 (1992);\\
A.~Khare, J.~Phys.\ A: Math.\ Gen.\ {\bf 25}, L749 (1992).
\bibitem{fsusy} C.~Ahn, D.~Bernard, and A.~Leclair, Nucl.\ Phys.~B {\bf 346}, 409 (1990);\\
L.~Baulieu and E.~G.~Floratos, Phys.\ Lett.~B {\bf 258}, 171 (1991);\\
R.~Kerner, J.~Math.\ Phys.\ {\bf 33}, 403 (1992);\\
S.~Durand, Phys.\ Lett.\ B {\bf 312}, 115 (1993);\\
S.~Durand, Mod.\ Phys.\ Lett.\ A {\bf 8}, 1795 (1993);\\
S.~Durand, Mod.\ Phys.\ Lett.\ A {\bf 8}, 2323 (1993);\\
A.~T.~Filippov, A.~P.~Isaev, and R.~D.~Kurdikov, Mod.\ Phys.\ Lett.~A {\bf 7},
2129 (1993);\\
N.~Mohammedi,  Mod.\ Phys.\ Lett.~A {\bf 10}, 1287 (1995);\\
N.~Fleury and M.~Rausch~de~Traubenberg, Mod.\ Phys.\ Lett.~A {\bf 11}, 2899 (1996);\\
J.~A.~de~Az\'carraga and A.~Macfarlane, J.\ Math.\ Phys. {\bf 37}, 1115 (1996);\\
R.~S.~Dunne, A.~Macfarlane, J.~A.~de~Az\'carraga, and J.~C.~P\'erez Bueno, Int.~J.~Mod.\ Phys.\ Lett.\ A {\bf 12}, 3275 (1997).
\end{thebibliography}
